\documentclass[fleqn,usenatbib]{mnras} 

\usepackage[T1]{fontenc}
\usepackage{ae,aecompl}

\usepackage{mathptmx}

\usepackage[dvips]{graphicx}
\usepackage{amsmath}                 
\usepackage{amssymb}                 
\usepackage{bm}                      

\def\bea{\begin{eqnarray}}
\def\eea{\end{eqnarray}}
\def\be{\begin{equation}}
\def\ee{\end{equation}}

\def\ms{M_\odot}
\def\mmax{M_\text{max}}
\def\fm3{$\text{fm}^{-3}$}

\def\mdu{M_\text{DU}}
\def\xdu{x_\text{DU}}
\def\rdu{\rho_\text{DU}}
\def\r1s0{\rho_{1S0}}
\def\m1s0{M_{1S0}}

\title
[Neutron star cooling with microscopic equations of state]
{Neutron star cooling with microscopic equations of state}

\author
[J.-B. Wei, G. F. Burgio, and H.-J. Schulze]
{J.-B. Wei, G. F. Burgio, and H.-J. Schulze
\\
INFN Sezione di Catania,
Dipartimento di Fisica e Astronomia, Universit\`a di Catania,
Via Santa Sofia 64, 95123 Catania, Italy
}

\date{\today}
\pubyear{2018}

\begin{document}
\label{firstpage}
\pagerange{\pageref{firstpage}--\pageref{lastpage}}
\maketitle

\begin{abstract}
We model neutron star cooling with several microscopic nuclear equations of state
based on different nucleon-nucleon interactions and three-body forces,
and compatible with the recent GW170817 neutron star merger event.
They all feature strong direct Urca processes.
We find that all models are able to describe well the current set of
cooling data for isolated neutron stars,
provided that
large and extended proton ${^1\!S_0}$ gaps and no neutron ${^3\!PF_2}$ gaps
are active in the stellar matter.
We then analyze the neutron star mass distributions
predicted by the different models and single out the preferred ones.
\end{abstract}

\begin{keywords}
stars: neutron --
dense matter --
equation of state.
\end{keywords}


\section{Introduction}

The recently observed first neutron star (NS) merger event GW170817
\citep{merger}
has allowed to substantially restrict the freedom of constructing NS
equations of state (EOS) compatible with observation.
In particular the determination of the NS tidal compressibility allowed
to derive severe constraints on the NS radii and the related stiffness of
the NS EOS \citep{radice,pascha,drago4,jinbiao}.
It turns out that microscopic EOSs compatible with those constraints
are rather stiff and feature fairly large proton fractions in the stellar matter,
which in turn implies the presence of the extremely strong Direct Urca (DU)
neutrino cooling process over an extended range of density in NS matter.

This allows to take a fresh look at the phenomenon of NS cooling,
where for a long time EOSs featuring DU cooling have been
disregarded in the so-called minimal cooling paradigm \citep{minimal},
which imposed a strong bias on the choice of the NS EOS.
In fact many microscopic nuclear EOSs do reach easily the required
proton fractions for the DU process \citep{zhl1,2010bs,zhl2,sel},
and we have already shown in \cite{2016MNRAS,2018MNRAS}
that a microscopic EOS featuring strong DU processes is very well able
to describe current cooling observations of isolated NSs
as well as the reheated cooling of the accreting NSs
in X-ray transients in quiescence \citep{yako14,Beznogov1,Beznogov2},
provided the DU process is dampened by the presence of nuclear pairing
throughout a sufficiently large (but not complete) set of the NS population.

The purpose of this work is to extend the
cooling simulations of \cite{2016MNRAS,2018MNRAS},
where only one specific microscopic EOS was used,
to three other microscopic EOSs that have also been identified as compatible
with the GW170817 event in \cite{drago4,jinbiao}.
All these EOSs fulfill upper and lower constraints on the tidal compressibility
derived from the interpretation of the merger event.
We investigate now which specific features of the EOSs determine the
NS cooling properties and compare the results obtained with the four EOSs.

This paper is organized as follows.
In Section~\ref{s:eos} we give a brief overview of the theoretical framework,
namely the Brueckner-Hartree-Fock (BHF) formalism adopted for the EOS,
the various cooling processes,
and the pairing gaps obtained with the same interactions.
Section~\ref{s:res} is devoted to the presentation and discussion
of the results for the cooling diagrams and the deduced NS mass distributions.
Conclusions are drawn in Section~\ref{s:end}.

\section{Formalism}
\label{s:eos}

\subsection{Equation of state}

One of the key ingredients of cooling simulations is the EOS.
In this paper we assume (without further justification)
the absence of exotic components like hyperons and/or quark matter,
such that the composition of the NS core is
asymmetric, charge-neutral, and beta-stable matter made of nucleons and leptons.

In our model, we calculate the EOS of nuclear matter within the BHF theoretical
approach \citep{1976Jeu,1999Book,2012Rep},
in which the starting point is the Brueckner-Bethe-Goldstone
equation for the in-medium $G$-matrix,
whose only input is the nucleon-nucleon (NN) bare potential $V$,
\be
 G[\rho;\omega] = V + \sum_{k_a k_b} V {{|k_a k_b\rangle Q \langle k_a k_b|}
 \over {\omega - e(k_a) - e(k_b)}} G[\rho;\omega] \:,
\label{e:g}
\ee
where
$\rho=\sum_{k<k_F}$ is the nucleon number density,
$\omega$ is the starting energy,
and the Pauli operator $Q$ determines the propagation of
intermediate baryon pairs.
The single-particle (s.p.) energy reads
($\hbar=c=1$)
\be
 e(k) = e(k;\rho) = {k^2\over 2m} + U(k;\rho) \:,
\label{e:en}
\ee
where the s.p.~potential $U(k;\rho)$ is calculated in the so-called
{\em continuous choice} and is given by
\be
 U(k;\rho) = {\rm Re} \sum_{k'<k_F}
 \big\langle k k'\big| G[\rho; e(k)+e(k')] \big| k k'\big\rangle_a \:,
\ee
where the subscript $a$ indicates antisymmetrization of the matrix element.
Finally the energy per nucleon is expressed by
\be
 {E \over A} =
 {3\over5}{k_F^2\over 2m} + {1\over{2\rho}} \sum_{k<k_F} U(k;\rho) \:.
\ee

In this scheme, we use several different nucleon-nucleon potentials
as bare NN interaction $V$ in the Bethe-Goldstone equation~(\ref{e:g}),
in particular the Argonne $V_{18}$ \citep{v18},
the Bonn B (BOB) \citep{1987PhR,Machleidt1989},
and the Nijmegen 93 (N93) \citep{1978PRD_Nagels,1994PRC_Stoks}.
These two-body forces are supplemented by suitable three-nucleon forces (TBF),
which are introduced in order to reproduce correctly
the nuclear matter saturation point.
In particular, BOB and N93 have been supplemented by microscopic TBF
employing the same meson-exchange parameters as the NN potential
\citep{1989Grange,2002Zuo,Li2008bp,zhl1},
whereas $V_{18}$ is combined either with the microscopic
or a phenomenological TBF,
the latter consisting of an attractive term due to two-pion exchange
with excitation of an intermediate $\Delta$ resonance,
and a repulsive phenomenological central term
\citep{Carlson:1983kq,Schiavilla:1985gb,1997Pud,1997A&A}.
They are labeled as V18 and UIX, respectively,
throughout the paper and in all figures.

Further important ingredients in the cooling simulations are the
neutron and proton effective masses,
\be
 \frac{m^*(k)}{m} = \frac{k}{m} \left[ \frac{d e(k)}{dk} \right]^{-1} \:,
\ee
which we derived consistently from the BHF s.p.~energy $e(k)$, Eq.~(\ref{e:en}),
see \cite{meff} for the numerical parametrizations.
Their effect is small compared to other uncertainties regarding
the cooling, and therefore in this paper we use the bare nucleon mass,
at variance with our previous simulations \citep{2016MNRAS},
where the effective masses were used in a consistent manner.

\subsection{Cooling processes}
\label{s:cp}

One important tool of analysis is the
temperature(or luminosity)-vs.-age cooling diagram,
in which currently a few ($\sim20$) observed isolated NSs are located.
NS cooling is over a vast domain of time
($10^{-10}-10^5\,\text{yr}$)
dominated by neutrino emission due to several microscopic processes
\citep{2001rep,2006ARNPS,2006PaWe,2007LatPra,potrev}.
The theoretical analysis of these reactions requires the knowledge of the
elementary matrix elements,
the relevant beta-stable nuclear EOS, and, very important,
the superfluid properties of the stellar matter, i.e.,
the gaps and critical temperatures in the different pairing channels.

The main contribution to the cooling comes from the
neutrons, protons, and electrons in the NS core.
In a non-superfluid NS, the most powerful neutrino process is the DU process,
which is in fact the neutron $\beta$-decay followed by its inverse reaction:
\be
 n \rightarrow p + e^-\! + \bar{\nu}_e
 \qquad \textrm{and} \qquad
 p+e^- \rightarrow n+\nu_e \:.
\label{e:DU}
\ee
However, the energy and momentum conservation imposes a density threshold
on this process \citep{1991LP}.
Various less efficient neutrino processes may be operating in the NS core
\citep{2001rep},
and dominate when the DU process is forbidden or strongly reduced.
The two main ones are the so-called modified Urca (MU) process:
\be
 n + N \rightarrow p + e^-\! + \bar{\nu}_e + N
 \qquad \textrm{and} \qquad
 p + e^-\! + N \rightarrow n + \nu_e + N \:,
\label{e:MU}
\ee
where $N$ is a spectator nucleon that ensures momentum conservation,
and the nucleon-nucleon bremsstrahlung:
\be
 N+N \rightarrow N+N+ \nu+\bar{\nu} \:,
\ee
with $N$ a nucleon and
$\nu$, $\bar{\nu}$ an (anti)neutrino of any flavor.

\begin{figure}
\vspace{-17mm}
\centerline{\hskip5mm\includegraphics[angle=0,scale=0.32]{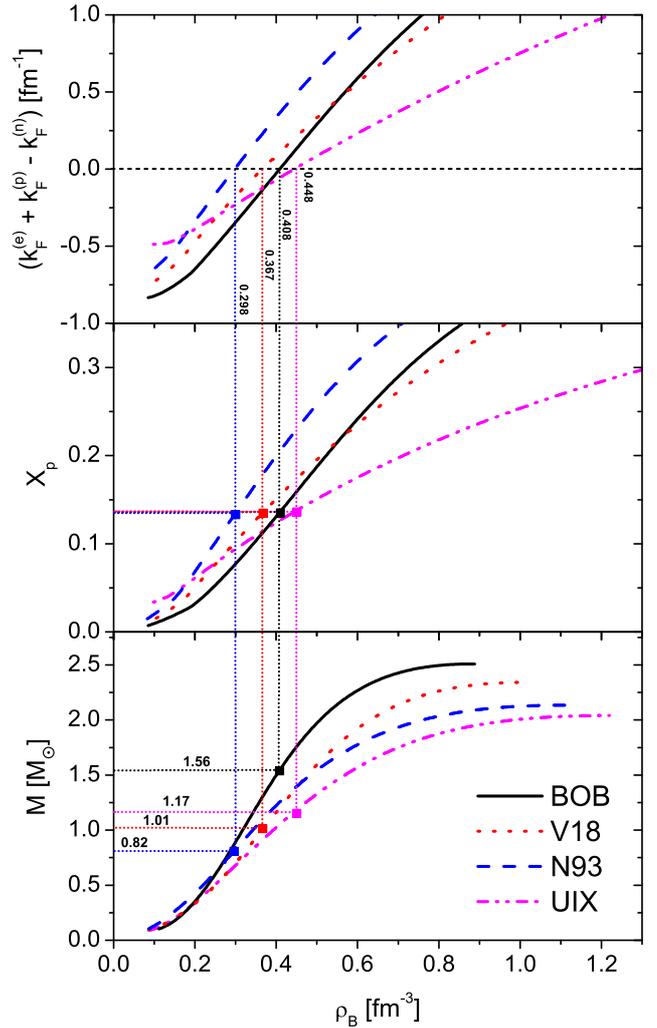}}
\vspace{-42mm}
\caption{
DU onset condition (upper panel),
proton fraction (central panel),
and NS mass (lower panel)
vs.~the (central) baryon density for the different EOSs.
The vertical dotted lines indicate the threshold density for the DU process.}
\label{f:mrho}
\end{figure}

For the different EOSs used in this work,
the DU process sets in at slightly different values of the proton fraction $x_p$
due to the presence of muons,
as shown in Fig.~\ref{f:mrho}.
The threshold values $\xdu$ are calculated starting from Eq.~(\ref{e:DU}),
in which the momentum conservation imposes the triangle rule, i.e.,
\be
 k_F^{(n)} < k_F^{(p)} + k_F^{(e)} \:.
\label{e:thres}
\ee
This is indicated by the vertical dotted lines in Fig.~\ref{f:mrho}
for the different EOSs,
and $\xdu$ is comprised in the range $0.133 < \xdu < 0.136$.
It is important to determine at which corresponding value $\rdu$
of the nucleon density
of beta-stable and charge-neutral matter the DU process sets in,
because compact stars characterized by central densities larger than
$\rdu$ will cool down very rapidly.
This is also displayed in Fig.~\ref{f:mrho}
and occurs in a density range between 0.30 and 0.45 \fm3
depending on the EOS.
In the lower panel we display the NS mass-central density relations
obtained by solving the Tolman-Oppenheimer-Volkoff equations
for hydrostatic equilibrium.
The NS masses $\mdu$ corresponding to the central densities $\rdu$
span a range between 0.82 (N93) and 1.56 (BOB) $\ms$,
above which the DU process can potentially operate.
We notice that in all cases the value of the maximum mass $\mmax$
is larger than the current observational lower limits
\citep{demorest2010,heavy2,fonseca16}.

The main results for $\xdu$, $\rdu$, $\mdu$, and $\mmax$
are also listed in Table~\ref{t:eos}.
We conclude that in all cases there is a wide range of NS masses
where the DU process is operative,
practically for all NSs with the V18 and N93 EOSs,
while only for the BOB EOS the threshold $\mdu=1.56\,\ms$ is high,
but in this case also $\mmax=2.51\,\ms$ is very large.

\def\myc#1{\multicolumn{1}{c}{$#1$}}
\begin{table}
\renewcommand{\arraystretch}{0.9}
\begin{center}
\caption{
Characteristic properties of several EOSs:
DU onset proton fraction $\xdu$, density $\rdu$, and
corresponding NS mass with that central density $\mdu$.
Upper limit of the range of p1S0 pairing $\r1s0$
and NS mass with that central density $\m1s0$.
Maximum NS mass $\mmax$.
Densities are given in \fm3 and masses in $\ms$.}
\label{t:eos}
\medskip
\begin{tabular}{@{}lrrrrrr}
\hline
  EOS & $\xdu$ & $\rdu$  & $\mdu$  & $\r1s0$ & $\m1s0$ & \myc{\mmax} \\
\hline\\[-3mm]
  BOB     & 0.1357 & 0.41 & 1.56 & 0.59 & 2.23 & 2.51  \\
  V18     & 0.1348 & 0.37 & 1.01 & 0.60 & 1.91 & 2.34  \\
  N93     & 0.1331 & 0.30 & 0.82 & 0.52 & 1.59 & 2.13  \\
  UIX     & 0.1363 & 0.45 & 1.17 & 0.70 & 1.70 & 2.04  \\
\hline
\end{tabular}
\end{center}
\end{table}

We finally remind the
possible strong constraints on NS cooling imposed by the
speculated very rapid cooling of the supernova remnant Cas~A
\citep{2009Nat,2010HeiHo,2013Elsha,2015HoPRC},
which we have studied in detail in \cite{2016MNRAS}.
As the observational claims remain highly debated \citep{casno1,casno2},
we do not consider this scenario in this work.

\subsection{Pairing gaps}
\label{s:gaps}

The effect of the neutron or proton superfluidity
in the dominant 1S0 and 3P2 channels
on the neutrino emissivity is twofold \citep{2001rep}.
On one hand, when the temperature decreases below the critical superfluid
temperature $T_c$ of a given type of baryons,
the neutrino emissivity of processes involving a superfluid baryon
is exponentially reduced,
together with the specific heat of that component.
For example, proton superfluidity in the core of a NS suppresses both Urca
processes but does not affect the neutron-neutron bremsstrahlung.
On the other hand, the pairing of baryons initiates a new type
of neutrino reactions called the pair breaking and formation (PBF) processes.
The energy is released in the form of a neutrino-antineutrino pair
when a Cooper pair of baryons is formed.
The process starts when $T=T_c$,
is maximally efficient when $T\approx0.8\,T_c$,
and is exponentially suppressed for $T\ll T_c$ (\citealt{2001rep}).
Therefore an essential ingredient for cooling simulations
is the knowledge of the pairing gaps for neutrons and protons
in beta-stable matter.

\begin{figure}
\vspace{-14mm}
\centerline{\includegraphics[angle=0,scale=0.32]{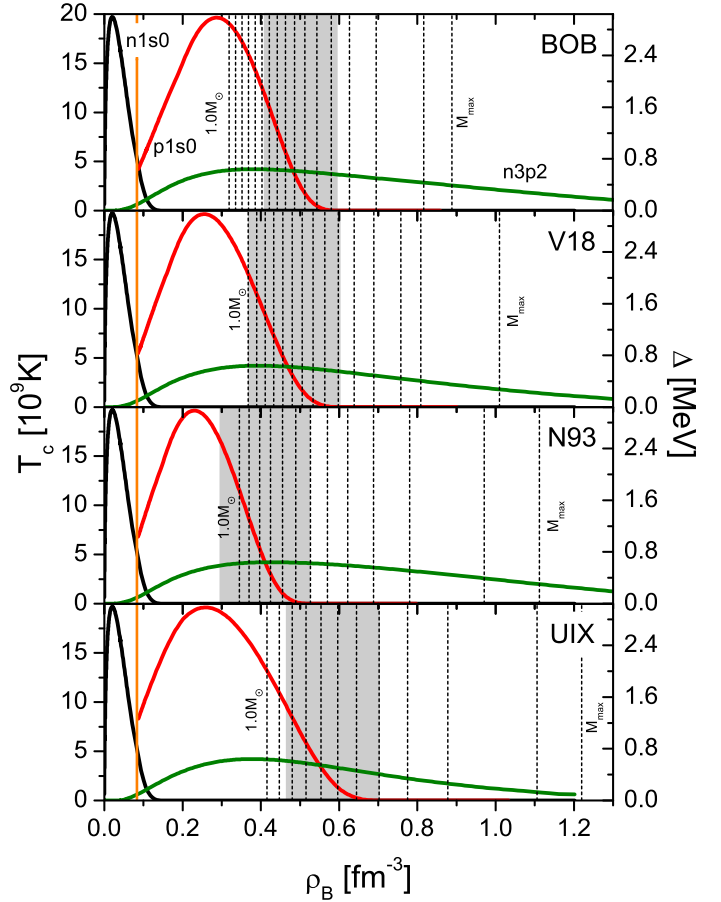}}
\vspace{-39mm}
\caption{
BCS gaps in the n1S0, p1S0, and n3P2 channels
in NS matter for the different EOSs.
The vertical dotted lines indicate the central density
of NSs with different masses
$M/M_\odot=1.0,1.1,\ldots$, up to the maximum mass value.
The shaded areas indicate the region between DU onset density $\rdu$
and vanishing of the p1S0 gap at $\r1s0$
(listed in Table~\ref{t:eos}),
i.e., where the DU process is blocked by pairing.
The orange vertical line represents the crust-core boundary.}
\label{f:gaps}
\end{figure}

In a previous paper \citep{2016MNRAS} we presented calculations obtained
using gaps computed consistently with the EOS, i.e.,
based on the same $V_{18}$ NN interaction and using the same medium effects
(TBF and effective masses), as shown in \cite{ourgaps}.
In this work we extend those studies by employing different nucleonic potentials.
Before illustrating the new results,
we remind the reader that the pairing gaps were computed on the BCS level
by solving the (angle-averaged) gap equation
\citep{bcsp1,bcsp2,bcsp3,bcsp4,bcsp5,bcsp6}
for the $L=0$ (1S0 gaps)
and the two-component $L=1,3$ (3P2 gaps) gap functions,
\be
 \left(\!\!\!\begin{array}{l} \Delta_1 \\ \Delta_3 \end{array}\!\!\!\right)\!(k)
 = - {1\over\pi} \int_0^{\infty}\!\! dk' {k'}^2 {1\over E(k')}
 \left(\!\!\!\begin{array}{ll}
 V_{11}\! & \!V_{13} \\ V_{31}\! & \!V_{33}
 \end{array}\!\!\!\right)\!(k,k')
 \left(\!\!\!\begin{array}{l} \Delta_1 \\ \Delta_3 \end{array}\!\!\!\right)\!(k')
\label{e:gap}
\ee
with
\be
  E(k)^2 = [e(k)-\mu]^2 + \Delta_1(k)^2 + \Delta_3(k)^2 \:,
\ee
while fixing the (neutron or proton) density,
\be
 \rho = {k_F^3\over 3\pi^2}
 = 2 \sum_k {1\over 2} \left[ 1 - { e(k)-\mu \over E(k)} \right] \:.
\label{e:rho}
\ee
Here $e(k)$ are the BHF s.p.~energies, Eq.~(\ref{e:en}),
containing contributions due to two-body and three-body forces,
$\mu \approx e(k_F)$ is the chemical potential
determined self-consistently from Eqs.~(\ref{e:gap}--\ref{e:rho}),
and
\be
 V^{}_{LL'}(k,k') =
 \int_0^\infty \! dr\, r^2\, j_{L'}(k'r)\, V^{TS}_{LL'}(r)\, j_L(kr)
\label{e:v}
\ee
are the relevant potential matrix elements
($T=1$ and
$S=0$; $L,L'=0$ for the 1S0 channel,
$S=1$; $L,L'=1,3$ for the 3P2 channel).
The relation between (angle-averaged) pairing gap at zero temperature
$\Delta \equiv \sqrt{\Delta_1^2(k_F)+\Delta_3^2(k_F)}$
obtained in this way and the critical temperature of superfluidity is then
$T_c \approx 0.567\Delta$.

In \cite{2016MNRAS,2018MNRAS}
we concluded that a very good description of cooling properties can be obtained
by just using the (eventually scaled) BCS values in the p1S0 channel and
disregarding any pairing in the n3P2 channel.
The complete suppression of the 3P2 gaps could be caused by
polarization corrections \citep{pol1,ppol1,ppol2,pol,pol3,ppol3},
which for the 1S0 channel are known to be repulsive,
but for the 3P2 are still essentially unknown;
and this might change the value of these gaps even by orders of magnitude.

For illustration we display in Fig.~\ref{f:gaps} the BCS pairing gaps
and critical temperatures
as a function of baryonic density of beta-stable matter for the different EOSs.
In this way one can easily identify which range of gaps is active
in different stars,
whose central densities are shown by vertical dotted lines for given NS masses.

Regarding the p1S0 pairing,
an important information is the density $\r1s0$ at which the
BCS gap disappears,
and the corresponding NS mass $\m1s0$ with that central density.
The DU process can only be completely blocked for NSs with $M<\m1s0$,
whereas for heavier stars it is active and unblocked in a certain domain
of the core, which leads to extremely fast cooling of these objects.
The value of $\m1s0$ is also listed in Table~\ref{t:eos}
and together with $\mdu$
determines the ranges of blocked and unblocked DU cooling.
We observe that for the BOB, V18, UIX, N93 EOS
the DU blocking terminates at $M/\ms=2.23,1.91,1.70,1.59$, respectively,
which implies very rapid cooling for heavier stars,
reflected in the following cooling diagrams.
The regions of blocked DU cooling are represented by shading
in Fig.~\ref{f:gaps}.

\begin{figure*}
\vspace{-40mm}
\centerline{\hskip-8mm\includegraphics[angle=0,scale=0.9,clip]{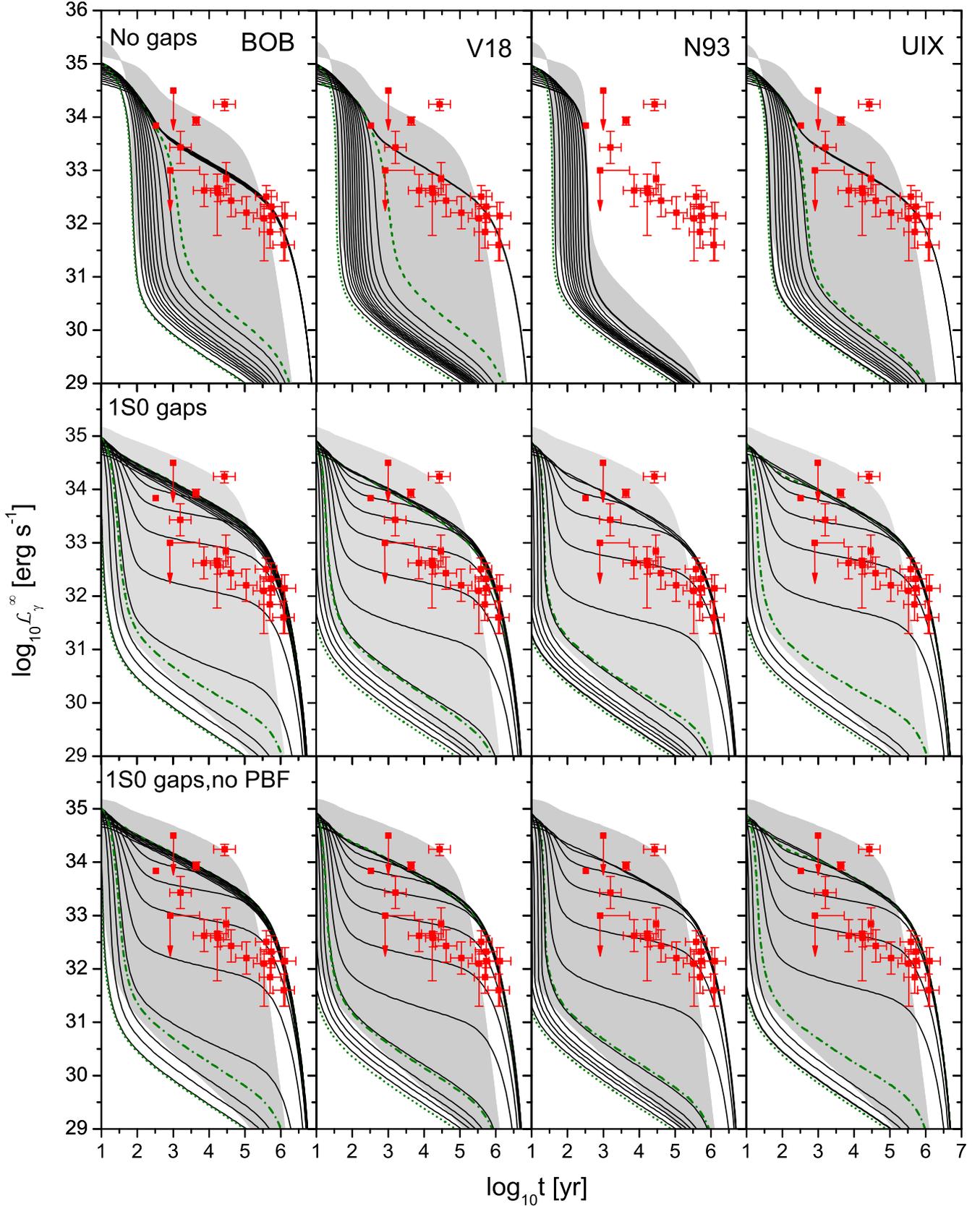}}
\vspace{-14mm}
\caption{
Cooling curves for different EOSs
without any pairing (upper panels),
with the inclusion of n1S0 and p1S0 BCS gaps (central panels),
and excluding the PBF processes in the latter case (lower panels)
for different NS masses $M/\ms=1.0,1.1,\ldots,\mmax$
(decreasing curves).
The dashed green curves mark the NS mass $\mdu+0.01\,\ms$
at which the DU process has just set in,
the dash-dotted green curves mark the NS mass $\m1s0$
for which the p1S0 gap vanishes in the center of the star,
and the dotted green curves correspond to $\mmax$ for each EOS.
The black curves are obtained with a Fe atmosphere and the shaded areas
cover the same results obtained with a light-elements
($\eta=10^{-7}$) atmosphere.
The data points are from \citep{Beznogov1}.
See text for more details.
}
\label{f:cool}
\end{figure*}

\section{Results}
\label{s:res}

Having quantitatively specified EOS and pairing gaps,
the NS cooling simulations are carried out using the
widely used code {\tt NSCool} \citep{Pageweb},
which comprises all relevant cooling reactions:
DU, MU, PBF, and Bremsstrahlung,
including modifications due to pairing discussed before.
In order to assess the uncertainty of the heat-blanketing effect
of the atmosphere,
we compare in the following results obtained with
a non-accreted heavy-elements (Fe) atmosphere 
and one containing also a maximum fraction
$\eta=\Delta M/M=10^{-7}$
of light elements from accretion, see \cite{potek}.

Our set of observational cooling data comprises the
(age, temperature) information
of the 19 isolated NS sources listed in \cite{Beznogov1},
where it was also pointed out that in many cases the distance to the object,
the composition of its atmosphere, thus its luminosity,
and its age are rather estimated than measured.
Thus in these cases, we use large error bars (a factor 0.5 and 2)
to reflect this uncertainty.

\subsection{Cooling diagrams}
\label{s:rescool}

For a better understanding,
we begin by discussing the simulations obtained with different EOSs
without including any superfluidity.
Results are displayed in Fig.~\ref{f:cool} (upper panels),
where the luminosity vs.~age is plotted for several NS masses
in the range $1.0,1.1,...,\mmax$.
The dashed green curves mark the NS mass $\mdu+0.01\ms$
at which the DU process has just set in,
whereas the dotted green curves correspond to the maximum mass $\mmax$.

The results are clearly unrealistic,
as observed NSs would essentially be divided into very hot ones and very cold
ones by the DU threshold $\mdu$,
with very few stars in between:
In a NS with $M<\mdu$, the DU process is turned off
and therefore the total neutrino emissivity
is orders of magnitude smaller than for a NS with a mass above the DU threshold.
Consequently the former has at a given age has a much higher luminosity
than the latter.
All NSs with $M<\mdu$ have a small neutrino emissivity,
hence their cooling curves are nearly indistinguishable on the scale
of Fig.~\ref{f:cool},
while for objects with $M>\mdu$,
the larger is the mass and thus the bigger is the central region of the star
where the DU process operates,
the lower is the luminosity and the cooling curves are no longer superimposed.

This feature depends on the EOS as explained in Sect.~\ref{s:cp}.
For example, in the N93 case the onset takes place at a very small value
of the density and the related gravitational mass $\mdu=0.82\,\ms$,
and therefore all NS masses undergo DU processes.
In the V18 case, the DU process starts for a 1.01\,$\ms$ NS,
and therefore only the first upper curve is influenced by MU alone,
whereas the remaining ones are determined by DU cooling.
The other EOSs have higher threshold values of
$\mdu/\ms=1.17,1.56$ for the UIX and BOB, respectively.

We now discuss the cooling curves switching on the n1S0 and p1S0 gaps,
as shown in Fig.~\ref{f:cool} (central panels).
For the gaps used in this work,
the main effect of superfluidity on NSs with $M>\mdu$
(dashed green curves, partially covered)
is the strong quenching of the DU process,
and thus a substantial reduction of the total neutrino emissivity.
Hence those stars have a higher luminosity compared to the non-superfluid case.
On the other hand, if $M>\m1s0$
(dash-dotted green curves),
the complete blocking of the DU process disappears
and the star cools very rapidly again.
One observes results in line with these features in the figure,
namely between $\mdu$ and $\m1s0$ there is now a smooth dependence of
the luminosity on the NS mass for a given age.
The effect is qualitatively the same for all EOSs,
just the distribution of NS masses in the cooling diagram depends on the EOS,
which will be analyzed in the next section.

Regarding the effect of the atmosphere models,
we note that by assuming a proper atmosphere for any given data point,
all current cooling data could potentially be explained in
the present scenario,
by assigning
a Fe atmosphere (black curves) to the oldest objects
and an accreted light-elements atmosphere (shaded area) to the hottest ones.
The currently known most extrem object,
XMMU J1731-347,
($\log_{10}{t}\approx4.4$, $\log_{10}{L_\gamma^\infty}\approx34.2$),
is indeed supposed to have a carbon atmosphere,
see the discussions in \cite{Beznogov1,Beznogov2,2018MNRAS}.

Although in general the presence of superfluidity is slowing down the cooling,
the PBF processes might prevail in certain situations and provide
an accelerated cooling of certain stellar configurations
\citep{2011PRLPage,2011MNRASSHTE,2011MNRASYA,2016MNRAS}.
It is therefore of interest to show in Fig.~\ref{f:cool} (lower panels)
also results where the 1S0 PBF processes have been switched off by hand.
It can be seen, however, that the effect of PBF cooling in the 1S0 channels
is practically negligible.

We conclude that the BCS p1S0 gap alone is able to suppress sufficiently
the DU cooling and to yield realistic cooling curves,
provided that it extends over a large enough density/mass range.
This is the case for all considered EOSs,
which yield however different mass profiles in the luminosity vs.~age plane.
This illustrates the necessity of precise information
on the masses of the NSs in the cooling diagram,
without which no theoretical cooling model can be verified.
We study this issue in some detail in the following.

\begin{figure}
\vskip-13mm
\centerline{\includegraphics[angle=0,scale=0.33,clip]{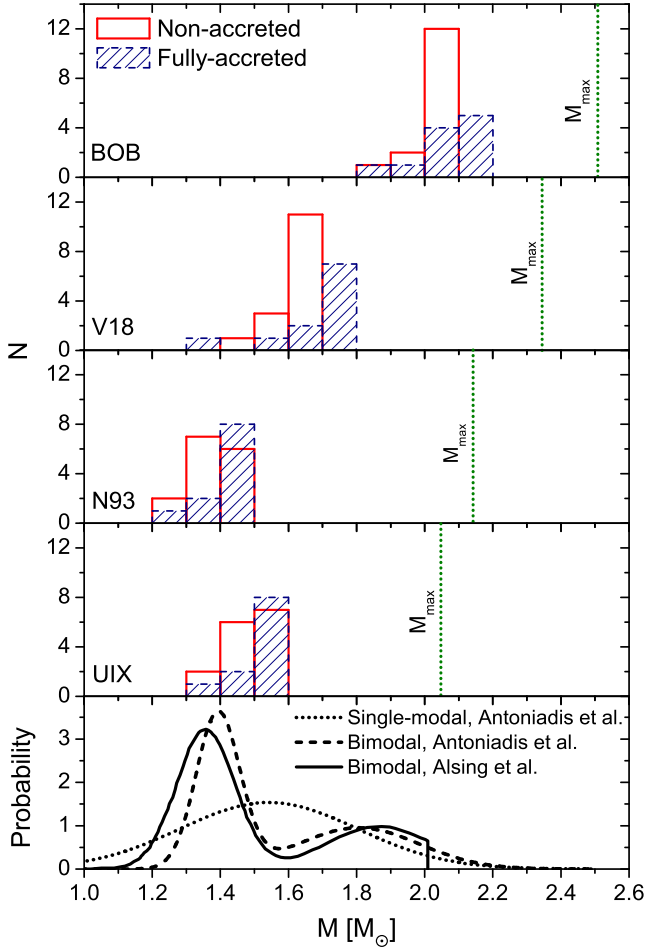}}
\vskip-13mm
\caption{
Deduced NS mass distributions from the cooling diagrams in
Fig.~\ref{f:cool} (central panels)
for the different EOSs.
Maximum masses are also indicated.
The lowest panel shows some recent theoretical results
\citep{anton16,alsing18}.
}
\label{f:md}
\end{figure}

\begin{figure}
\vskip-4mm
\centerline{\includegraphics[angle=0,scale=0.31,clip]{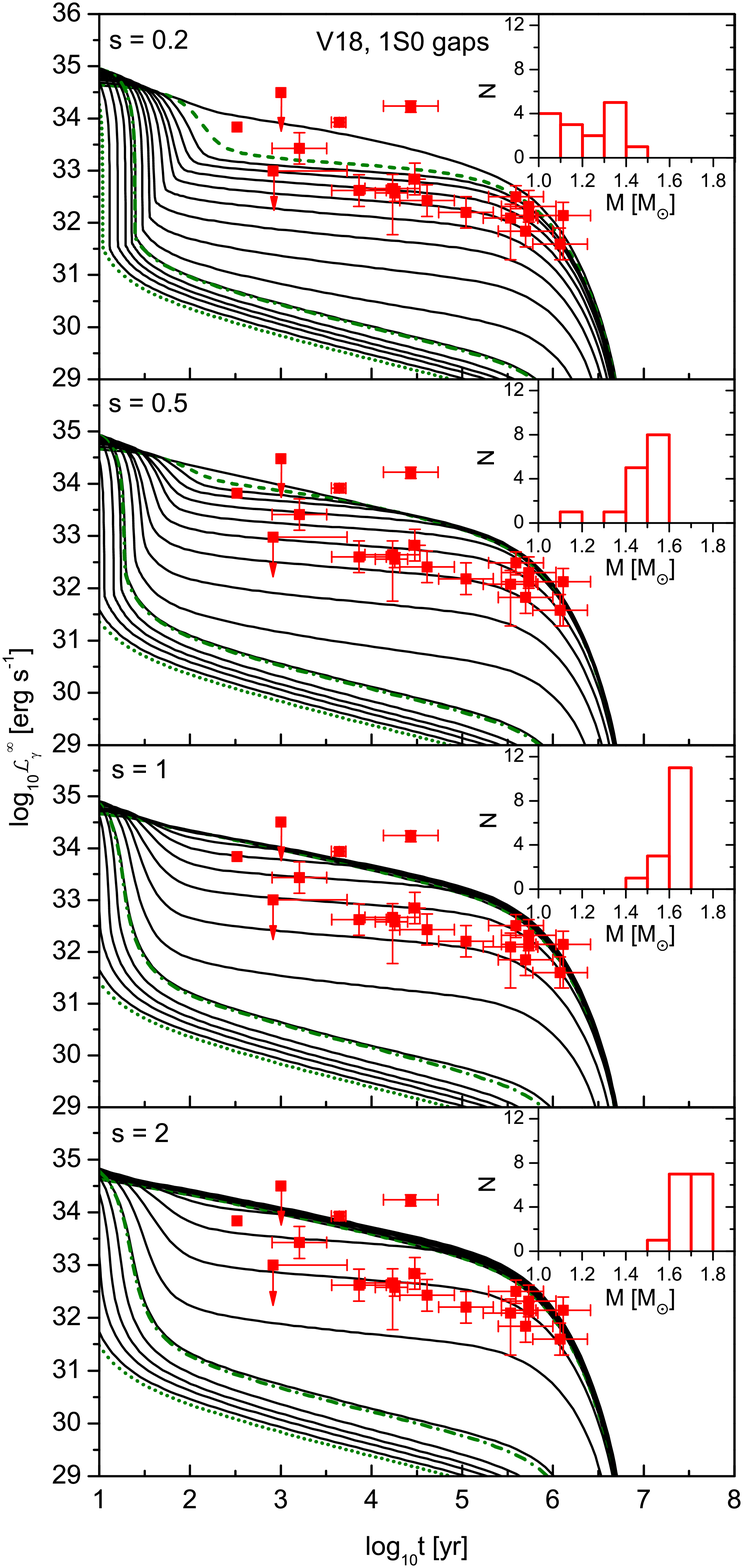}}
\vskip-9mm
\caption{
Cooling curves obtained with the V18 EOS, Fe atmosphere,
and 1S0 BCS gaps scaled by
factors $s=0.2,0.5,1.0,2.0$.
The insets show the derived mass distributions.
}
\label{f:scale}
\end{figure}

\subsection{Neutron star mass distributions}

Assuming that the currently observed set of isolated NSs in the cooling
diagrams of Fig.~\ref{f:cool}
reflects the unbiased mass distribution of NSs in the Universe
(which is highly unlikely due to strong selection effects;
for example, very old and massive NSs are too faint to be observed
and would therefore never appear in the cooling diagrams.
Also, the mass distribution of isolated NSs could be very different from those
in binary systems, etc.),
one can extract straightforwardly the predicted mass distribution
of NSs from the figures.
For simplicity we disregard the error bars in this analysis.
The results are shown as histograms in Fig.~\ref{f:md},
obtained directly from the binning in mass intervals in Fig.~\ref{f:cool}.
One observes clear differences between the four EOSs,
with surprisingly small dependence on the atmosphere model.
The lowest panel provides for comparison
a compilation of recent theoretical results
for the NS mass distribution \citep{ozel12,kizil13,anton16,alsing18}.
We stress again, however, that there is no good reason that the
mass distribution extracted from the cooling data in this way
should be similar to the overall mass distribution of NSs in the Universe
or even to that of isolated NSs only.

Due to this problem and the scarcity of data,
no firm conclusions can be drawn for the moment,
apart from perhaps excluding the BOB model,
which does not predict any mass even close to the NS canonical value
of about $1.4\,\ms$.
This model features also a very large $\mmax=2.51\,\ms$,
which seems to be in conflict with recent upper limits on $\mmax$
derived from analysis of the NS merger event \citep{shiba17,marga17,rezz18}.
On the other hand,
the $\mmax=2.04\,\ms$ of the UIX EOS appears too small,
which leaves as most realistic models either N93 or V18.
Clearly more data points, ideally with assigned known masses,
would be required for a more profound analysis of this kind.

To emphasize even more the value of data with well-assigned masses,
we point out that the mass distributions do not only depend on the EOS,
but also on the pairing gaps.
For that purpose we plot in Fig.~\ref{f:scale}
the results obtained with the V18 model and applying
different scaling factors $s = 0.2, 0.5, 1.0, 2.0$ to the 1S0 BCS gaps,
which could be motivated by the polarization effects discussed
in Sec.~\ref{s:gaps}.
One sees that while the overall coverage of the luminosity vs.~age plane
remains nearly unaffected,
the deduced NS mass distributions
(shown as insets)
depend sensitively on the gap scaling factor.
In the specific case,
one would be able to exclude very large scaling factors,
which is indeed physically reasonable.
Of course, modifying also the density domain of the pairing,
similar variations would be obtained in line with the considerations
in Sec.~\ref{s:rescool}.
But we think it is premature to try to resolve this issue
with the present set of cooling data.

\section{Conclusions}
\label{s:end}

Motivated by the recent availability of more stringent restrictions
on the NS EOS provided by a NS merger event,
we have analyzed in this work the predictions of some compatible
microscopic BHF EOSs for the cooling properties of isolated NSs.
All EOSs feature strong DU cooling for a wide range of masses
and the presence of superfluidity is required for realistic cooling scenarios.

We find that assuming absence of n3P2 pairing and employing n1S0 and p1S0 BCS
gaps with possible rather generous scaling factors,
a reproduction of all current cooling data for isolated NSs
can be achieved with any of the proposed EOSs.
A naive and straightforward analysis of the deduced NS mass distribution
would exclude only the stiffest BOB EOS,
which also predicts a fairly large maximum mass.

The combined and consistent analysis of different aspects of NS physics
like mergers, radius measurements, and cooling
will allow in the future an always more accurate derivation of the nuclear EOS,
also in view of the possible presence of exotic components like quarks
or hyperons,
which was excluded from the start in the present work,
but will be addressed in the future.

\section*{Acknowledgments}

We acknowledge helpful discussions with M.~Fortin
and financial support from ``PHAROS,'' COST Action CA16214,
and the China Scholarship Council
(CSC File No.~201706410092).

\bibliographystyle{mnras}     
\bibliography{coolmic}        
\bsp                          
\label{lastpage}
\end{document}